# Curating Model Problems for Software Designing

Mary Shaw
School of Computer Science
Carnegie Mellon University
Pittsburgh PA, USA
mary.shaw@cs.cmu.edu

Marian Petre
School of Computing and Communications
Open University
Milton Keynes, UK
m.petre@open.ac.uk

*Abstract*—Many disciplines use standard examples for education and to share and compare research results. The examples are rich enough to study from multiple points of view; they are often called model problems. Software design lacks such a community resource. We propose an activity for *Designing 2025* in which participants improve some existing model problem descriptions and initiate new ones – with a focus on use in software design education, plus potential utility in research.

## I. Model Problems in Education and Research

It is common for a discipline, especially one that is just getting its wits about itself, to adopt some shared, well-defined problems for teaching and research [1]. Known variously as model problems, model systems, or type problems, they provide a way to compare methods and results, work out new techniques on standard examples, and set a minimum standard of capability for new participants [2][3][4][5]. In time, a reasonable approach to some of these problems becomes the 'price of admission' for a new technique to get serious consideration. Model problems also provide pre-debugged sources of educational exercises, and they can be a vehicle for getting research ideas into education.

Biology, for example, has:
- *Drosophila melanogaster* (the fruit fly)
- *Rattus rattus Norvegicus* (the lab rat)
- *Escherichia coli* (the digestive bacterium)
- *Tobamovirus tabaci* (the tobacco mosaic virus) [5]

Each of these is part of the common language of discourse in its field. Each provides a familiar concrete instance that illustrates an important set of issues. This allows discussions to start from shared knowledge of the basic example and proceed expeditiously to the result, theory, or technique of current interest. It also facilitates comparison of research results.

Closer to home, computer science has model problems in many areas. Many of these originated in research and have shown their broad value as standard educational examples. Familiar examples include:
- *Algorithms and Data Structures:* Sort, search, greatest common divisor, prime integers, set, stack, queue
- *Synchronization:* Reader/writer, producer/consumer, dining philosophers, cigarette smokers
- *Formal Specifications:* Telegraph, lift (elevator, on the west side of the Atlantic), library
- S*oftware architecture:* Cruise control, meeting
- *Graphics:* Utah teapot, Stanford bunny
- *Combinatoric Optimization:* Travelling salesman

## II. Model Problems for Software Designing

Model problems for software design research and education should help us focus on specific design issues and be useful for illustrating and prompting the processes, reasoning, and dialogues of designing. Such issues include (for example):
- Representing different design alternatives, comparing them, and selecting among them.
- Engaging with different user / stakeholder perspectives that have implications for design.
- Describing and distinguishing among different kinds of system organization -- not only structural differences, but the implications of those differences.
- Providing opportunity to evaluate or test properties of designs.
- Engaging with omissions, slips, misconceptions, miscommunications, etc.

Each model problem should be rich enough to tackle from multiple design perspectives and show how different perspectives interact. Of course, no single such problem will serve all needs, so a rich collection is appropriate—but multiple perspectives on a smaller collection will be richer than a long list of problems that are only treated in a couple of ways.

The workshop will aim to address a few key issues: properties of good model problems, experiences of using model problems in education (e.g., presenting the problems in a way that promotes educational experiences; addressing challenges experienced in teaching designing), what it means to teach designing. Hence the aspiration is that the conversations both contribute to a repository and also contribute to broader discussions about what we should be teaching about designing —beyond presenting examples of design to engage students in the practice of designing.

## III. What Makes a Good Model Problem for Software Designing?

We look forward to discussion of the properties of a good model problem. Here is a starting set of candidate criteria:
- Presents a simple statement of a relatable problem – one that's explainable and explorable
- Can be explored at different levels of abstraction
- Raises high-level design issues: about the structure of the problem; the interaction between decisions; and also about the distinction between the world and the machine

Supported in part by the Alan J. Perlis Professorship in Computer Science at Carnegie Mellon University.

- Does not have a single "right" answer, but exposes tradeoffs that depend on requirements, context, problem framing, constraints, stakeholder values

Good model problems can serve both research and education. A collection of model problems is different from a collection of case studies [6] or a collection of failures for analysis. These tend to be analyses of specific designs and implementations.

We and others have started a repository for design model problems [7]. It currently contains a few problem statements and examples of how they have been used in education or research (e.g., Cruise Control, Traffic Signal Simulation, London Ambulance, BikePark). We would like to make this a community resource, and to that end we invite workshop participants to contribute.

## IV. Workshop Activity

The purposes of this activity are to use the model problems as a vehicle for broader discussion about teaching designing, improve the nascent model problem repository [7], and build a community of colleagues who will use and improve the material through time.

### A. Advance Preparation

The workshop activity will be promoted in advance to workshop registrants, to give them time to identify/prepare material (e.g., an example model problem, or the use of a problem from elsewhere) from their own teaching and/or research to share in the discussions.

### B. Workshop Activity

Depending on the facilities and number of participants overall, participants will divide into smaller groups (max. 8-10). Groups may be organised around model problems or self-forming, depending on the participants. The time (1-1/2 hours) will be divided into a short introduction, two discussion periods with a break in between, and collective feedback. We shall identify a strong rapporteur in each group (who will record notes in a shared document), and the participants and groups will be encouraged to sketch and make notes throughout the discussion (flip charts, post-its, the shared document, etc.).

The first session will discuss the participants' **experience of teaching software designing** using example problems, based on priming questions**:**

- What are participants' experiences using examples or model problems in teaching designing? How are design problems selected? How are they used? What are the challenges participants have faced in teaching designing? What are the gaps? What are the outcomes?
- What does it mean to teach software designing (rather than just sharing designs)? What should we be teaching? How do educators shape formative activities using model problems (or otherwise)? What are the design issues they address?
- How do participants find examples for use in teaching? What are the properties of good model problems for educational use?

The second session will build on this by focusing on **what makes a good model problem for software designing**. It will begin by introducing any model problems that participants bring to the table. The group is invited to discuss the problems, how they're characterized, how they've been used (or might be used), what concepts or design activities they're useful for, etc—with the aim of refining the list of criteria (Section III).

Each group will be given latitude to decide what it wants to prioritize. The workshop is fundamentally about discussion, so we expect to be flexible in the execution.

### C. Post-Workshop Followup: Model Problem Repository

Participants will be asked to reflect after the workshop, to add material to the repository [7], and to suggest improvements or follow-up discussions, etc. Reflection may include consideration of the use of model problems for both teaching and research. The repository improvements may be the addition of new problems; new perspectives on the use/contextualization of the problems in teaching (e.g., usage to address different learning goals); or improvements of the presentation.

The repository uses a standard template to give a full description of the problem, to list design issues and tradeoffs related to the problem, to include or link to ways the problem has been analyzed or otherwise used, all with attributions.

We shall identify a curator for the repository, to address questions, identify opportunities, and provide light-touch moderation. The intention is for the repository to be a community effort, hence accessible, hosting multiple file types, editable by lots of people, etc.

## Acknowledgments

Our thanks to Andre van der Hoek for his formative input, and to all who have contributed to the current repository.